\documentclass{aa}  

\usepackage{graphicx}
\usepackage{txfonts}
\usepackage{subfigure}
\begin{document} 

    \title{Multi-frequency point source detection with fully convolutional networks: Performance in realistic microwave sky simulations}
    \titlerunning{MultiPoSeIDoN in Realistic Simulations}
    \authorrunning{Casas J.M. et al.}
   
    \author{Casas J. M.\inst{1,2},
    Gonz{\'a}lez-Nuevo J.\inst{1,2},
    Bonavera L. \inst{1,2},
    Herranz D. \inst{4,5},
    Suarez Gomez S. L.\inst{2,3},
    Cueli M. M.\inst{1,2},
    Crespo D. \inst{1,2},
    Santos J. D. \inst{1,2},
    S{\'a}nchez M. L. \inst{1,2},
    S{\'a}nchez-Lasheras F. \inst{2,3},
    de Cos F. J. \inst{2,6}
}

   \institute{$^1$Departamento de F{\'i}sica, Universidad de Oviedo, C. Federico Garc{\'i}a Lorca 18, 33007 Oviedo, Spain\\
             $^2$Instituto Universitario de Ciencias y Tecnolog{\'i}as Espaciales de Asturias (ICTEA), C. Independencia 13, 33004 Oviedo, Spain\\
             $^3$Departamento de Matematicas, Universidad de Oviedo, C. Federico Garcia Lorca 18, 33007 Oviedo, Spain\\
             $^4$Departamento de Física Moderna, Universidad de Cantabria, 39005 Santander, Spain\\
             $^5$Instituto de Física de Cantabria, CSIC-UC, Av. de Los Castros s/n, 39005 Santander, Spain\\
              $^6$Escuela de Ingeniería de Minas, Energía y Materiales Independencia 13, 33004 Oviedo, Spain
             }


 \abstract{}{}{}{}{} 
  \abstract
   {Point source (PS) detection is an important issue for future cosmic microwave background (CMB) experiments since they are one of the main contaminants to the recovery of CMB signal on small scales. Improving its multi-frequency detection would allow us to take into account valuable information otherwise neglected when extracting PS using a channel-by-channel approach.}
   {We aim to develop an artificial intelligence method based on fully convolutional neural networks to detect PS in multi-frequency realistic simulations and compare its performance against one of the most popular multi-frequency PS detection methods, the matrix filters. The frequencies used in our analysis are 143, 217, and 353 GHz, and we imposed a Galactic cut of $30^\circ$.}
   {We produced multi-frequency realistic simulations of the sky by adding contaminating signals to the PS maps as the CMB, the cosmic infrared background, the Galactic thermal emission, the thermal Sunyaev-Zel'dovich effect, and the instrumental and PS shot noises. These simulations were used to train two neural networks called flat and spectral MultiPoSeIDoNs. The first one considers PS with a flat spectrum, and the second one is more realistic and general because it takes into account the spectral behaviour of the PS. Then, we compared the performance on reliability, completeness, and flux density estimation accuracy for both MultiPoSeIDoNs and the matrix filters.}
   {Using a flux detection limit of 60 mJy, MultiPoSeIDoN successfully recovered PS reaching the 90\% completeness level at 58 mJy for the flat case, and at 79, 71, and 60 mJy for the spectral case at 143, 217, and 353 GHz, respectively. The matrix filters reach the 90\% completeness level at 84, 79, and 123 mJy. To reduce the number of spurious sources, we used a safer 4$\sigma$ flux density detection limit for the matrix filters, the same as was used in the \textit{\emph{\emph{Planck}}} catalogues, obtaining the 90\% of completeness level at 113, 92, and 398 mJy. In all cases, MultiPoSeIDoN obtains a much lower number of spurious sources with respect to the filtering method.
   The recovering of the flux density of the detections, attending to the results on photometry, is better for the neural networks, which have a relative error of 10\% above 100 mJy for the three frequencies, while the filter obtains a 10\% relative error above 150 mJy for 143 and 217 GHz, and above 200 mJy for 353 GHz.}
   {Based on the results, neural networks are the perfect candidates to substitute filtering methods to detect multi-frequency PS in future CMB experiments. Moreover, we show that a multi-frequency approach can detect sources with higher accuracy than single-frequency approaches also based on neural networks.}


   \maketitle
%

\section{Introduction}

The search for point sources (PS) as contaminants to the recovery of the cosmic microwave background (CMB) anisotropies at small angular scales has become more and more relevant through the years since the conception of the Wilkinson Microwave Anisotropy Probe (WMAP, \citealt{BEN13}) and \textit{\emph{\emph{Planck}}} \citep{PLA18_IV} missions. 
At millimetre wavelengths, most of the PS are dusty galaxies, mainly dusty star-forming galaxies among which there are the most intense stellar nurseries in the Universe, and blazars, that is
active galactic nuclei (AGNs), whose jets are aligned in the line of sight of the satellite instrument.
In this regime, they are one of the main contaminants to the recovery of the CMB anisotropies on small scales. 
The future CMB experiments, such as the Probe of Inflation and Cosmic Origins \citep[PICO,][]{Han19}, the CMB-S4 \citep{CMBS4} and the Simons Observatory \citep[SO,][]{SO}, all of them with higher resolution than \textit{\emph{\emph{Planck}}}, are designed to keep the PS contamination low. This can be achieved by measuring at frequencies with lower PS contribution, by taking the data in regions with robust extragalactic surveys available, and mostly by developing high-performance methods for PS detection.

Multi-frequency detection of PS is an important field of research because the \textit{\emph{\emph{Planck}}} successors will be able to observe the sky at simultaneous wavelengths. The diffuse component separation is possible with multi-wavelength information, but the PS detection is not because they can have different physical properties, that is, each individual galaxy acting like a PS has its own unique spectral behaviour. Therefore, the PS detection is generally a task for single-frequency methods, and the catalogues of extragalactic sources are extracted from CMB maps one channel at a time. However, using this approach, valuable information that multi-wavelength experiments can offer is wasted. Thus, it is appropriate to develop accurate multi-frequency PS detection methods for future CMB experiments.

The first multi-frequency compact source detection methods made use of prior knowledge of the spatial profile of the sources. Moreover, they needed to know that sources should appear as compact objects in diffuse random fields, having the source located in a noisy map in order to help the detection. Some of those methods were wavelet techniques (\citealt{Vie01_2}, \citeyear{VIE03}; \citealt{GN06}; \citealt{Sanz06}; \citealt{LOP07}), Bayesian approaches (\citealt{Hobson03}; \citealt{Carvalho09}), linear filters (\citealt{Sanz01}; \citealt{Chiang02}; \citealt{Her02}, \citeyear{Herranz02_2}; \citealt{LOP04}, \citeyear{LOP05}, \citeyear{LOP05_2}), and matched filters (\citealt{TEG98}; \citealt{BAR03}; \citealt{LOP06}).

An evolution of these methods was based on combining simulated multi-wavelength maps to obtain a higher average in the signal-to-noise ratio of the sources. This approach was used in \citet{Naselsky02}. Similarly, \citet{Chen08} combined \textit{WMAP} \textit{W} and \textit{V} bands, obtaining CMB-free maps that helped them to detect the fainter sources. Although these methods detected a higher number of sources, they failed to recover their flux density.

The matrix filters (MTXFs; \citealt{Herranz08}) were introduced as an intermediate approach of the above mentioned methods. It looked for the detection of compact sources using their distinctive spatial behaviour and multi-wavelength information, but without the need to know their frequency dependence. After being applied to realistic microwave sky LFI \textit{\emph{\emph{Planck}}} simulations \citep{Herranz09}, MTXFs were used in \citet{PCCS} to validate the LFI channels of the \textit{\emph{\emph{Planck}}} catalogue of compact sources (PCCS). After that, they were used in \citet{PIPLIV} to produce the \textit{\emph{\emph{Planck}}} multi-frequency catalogue of non-thermal (i.e. synchrotron-dominated) sources (PCNT), which was the first multi-frequency catalogue of compact sources covering the nine \textit{\emph{\emph{Planck}}} channels.

In the last years, machine learning techniques have represented a revolution in astrophysics and cosmology. One of the most popular machine learning models concerns neural networks. These are artificial intelligence techniques inspired by the human brain, and they can be trained to learn non-linear behaviours from data by supervised or unsupervised learning (\citealt{SUA20}, \citeyear{SUA21}; \citealt{RIE19}, \citeyear{RIE20}). These characteristics make neural networks a good candidate to obtain interesting results in cosmology. Some examples of recent applications of neural networks in this field are the identification of galaxy mergers \citep{Pea19} and strong gravitational lenses (\citealt{Pet17}; \citealt{HEZ17}) in astronomical images, a galaxy classifier \citep{KIM17}, a better estimation of cosmological constraints from weak lensing maps \citep{Flu19}, a real-time gravitational wave detector \citep{GEO18}, a high-fidelity generation of weak lensing convergence maps \citep{Mus19}, and cosmological structure formation simulations under different assumptions \citep{Mat18, He19, Per19, Gui19}.

Multi-layer perceptron \citep{JUE12} and convolutional neural networks (CNNs) (\citealt{LeC89}, \citeyear{LEC15}) have been successfully applied to image processing (and related fields) for modelling and forecasting \citep{GRA13, GIU13}. An evolution of CNNs are the fully-convolutional neural networks (FCN) \citep{LON15, DAI16}, which are usually applied in image segmentation to classify a labelled object pixel-at-pixel by making use of different layers to learn image patterns (e.g. shapes, smoothness, and borders). 
In order to obtain relevant features of the input images, convolution and merging layers are generally paired in convolution and deconvolution steps. After that, the output (image or numerical) is obtained.

Fully-convolutional networks
were recently used to detect PS in noisy microwave background simulations \citep{BON21}. The neural network, called PoSeIDoN, was trained at 217 GHz, and its performance was compared at 143, 217, and 353 GHz against the Mexican hat wavelet 2 (MHW2; \citealt{GN06}), a single-frequency filtering method that was used in the \textit{\emph{\emph{Planck}}} experiment to detect PS at both LFI and HFI channels, creating the \textit{\emph{\emph{Planck}}} catalogue of compact sources \citep{PCCS} and the second \textit{\emph{\emph{Planck}}} catalogue of compact sources \citep{PCCS2}. PoSeIDoN obtained similar results on completeness with respect to the filter but with a much lower number of spurious detections.

In this work, we propose the use of FCN as a multi-frequency generalisation of PoSeIDoN to develop the Multifrequency Point Source Image Detection Network (MultiPoSeIDoN) 
to detect compact sources in noisy background maps by using image segmentation. 

The outline of the paper is the following. Section \ref{sec:simulations} covers how the simulated maps are generated. Section \ref{sec:methodology} describes our methodology. The results are explained in Section \ref{sec:results}, and our conclusions are discussed in Section \ref{sec:conclusions}.


\section{Simulations}
\label{sec:simulations}

In this work, realistic simulated maps of the microwave sky are used. These maps correspond to sky patches at the central channels of the \textit{\emph{\emph{Planck}}} mission: 143, 217, and 353 GHz. The pixel size is 90 arcsec, which is a round number close to the 1.72 arcmin used in the  \textit{\emph{\emph{Planck}}} maps, corresponding to $N_{\rm side}=2048$ in the \texttt{HEALPix} all sky pixelisation schema \citep{GOR05}. Balancing between the density of bright sources per patch and size, we consider that patches of 128$\times$128 pixels are sufficient for our study.

The multi-frequency PS detection is studied with two different sets of simulations. The first one consists of realistic simulated maps whose PS contribution is from flat-spectrum sources, that is, objects that have the same spectral behaviour in the three channels. The other set of simulations have PS with different spectral behaviours, thus also including in this case sources that increase and decrease their emission with frequency.

In the first case, where only sources with a flat spectrum have been considered, radio galaxies, mainly radio quasars and BL lacertae objects (or simply BL Lacs), commonly known as blazars, are simulated at 217 GHz using the model by \cite{TUC11} and the software CORRSKY \citep{GN05}. Then, in order to produce the simulated catalogue at 143, 217, and 353 GHz, the flux densities are convolved with the corresponding instrumental FWHM (7.22, 4.90, and 4.92 arcmin at 143, 217, and 353 GHz, respectively \citep{PLA18_IV}). 
Although not substantially contributing to the statistical properties of the background, the infrared late-type galaxies (IRLT, mainly starburst and local spiral galaxies; \citealt{Tof98}) are also simulated in order to add their shot-noise contribution to the patches by adopting the source number counts by \cite{CAI13}, normalised to the later update by \cite{Neg13} and the software CORRSKY. 

In order to obtain the set of simulations for the second case, the spectral behaviour of radio and IRLT sources is assumed to vary as
\begin{equation}
    S \hspace{2pt} = \hspace{2pt} S_{0} \hspace{1pt} \left( \frac{\nu}{\nu_{0}} \right)^{\alpha},
\label{eq:flux_density}
\end{equation}
where $S$ is the flux density of the sources at each frequency $\nu$, $S_{0}$ is the flux density at the central channel with frequency $\nu_{0}$, that is, 217 GHz, and $\alpha$ is the spectral index.

The spectral index distribution for each population is estimated in \textit{\emph{\emph{Planck}}} at each frequency \citep{PCCS2}. For frequencies below 217 GHz, the emission is mostly due to radio galaxies. Above such a frequency, most of the contribution comes from the late-type galaxies. At 217 GHz, the emission of both kind of sources is relevant to the total distribution.
In our simulations, the spectral index is randomly chosen according to the gaussian distributions with mean and standard deviation given in \cite{PCCS2}.

To obtain realistic simulations at these frequencies, we also consider the cosmic infrared background (CIB, \citealt{Pug96}; \citealt{Hau01}; \citealt{Dol06}) emission due to high redshift infrared PS (massive proto-spheroidal galaxies in the process of forming most of their stellar mass, \citealt{Gra04}; \citealt{Lap06}, \citeyear{LAP11}; \citealt{CAI13}), which is dominant at a few arcminute resolution, resulting in a contaminant for the purpose of our work.
To simulate them, the source number counts given by \cite{CAI13}, their angular power spectrum given by \cite{LAP11}, and the software CORRSKY are used. Their spectral behaviour is simulated by assigning them the same spectral index as for the late-type galaxies. 
For all the source populations (radio, late-type, and CIB) and frequencies, we simulate them down to the same flux density limit of 3 $\mu$Jy.

On larger angular scales, the contamination due to diffuse emission by our Galaxy and the CMB are also considered. Galactic emission varies significantly with frequency. Since our simulations correspond to the 143, 217, and 353 GHz \textit{\emph{\emph{Planck}}} channels, only thermal dust emission \citep{Finkbeiner1999} from our Galaxy is simulated. Both CMB and thermal dust emissions are introduced in the simulated maps by randomly selecting patches at the same position where the sources were simulated. On one hand, the CMB maps are the ones given at each frequency by the SEVEM method (\citealt{MartinezGonzalez2003}; \citealt{LEA08}; \citealt{FER12}). These maps are the inpainted version to fill the empty pixels due to the mask (both for the galactic plane and the PS) used during the component separation process \citep[see][and references therein]{PLA18_IV}. On the other hand, the thermal dust emission is added at each frequency by using the \textit{\emph{\emph{Planck}}} Legacy Archive (PLA) simulations, which were produced using the Planck Sky Model software \citep{Del13}. 

The \textit{\emph{\emph{Planck}}} maps are at $N_{\rm side}=2048$, which corresponds to a pixel size of 1.72 arcmin. The selected sky patches, which are randomly chosen to be negligible the probability of having two exact sky maps, are projected into flat patches of pixel size of 1.5 arcmin by using the \texttt{gnomview} function of \texttt{HEALPix} framework \citep{GOR05}. Small-scale fluctuations are added to Galaxy emission by following the \cite{Miv07} method in order to simulate the sky at \textit{\emph{\emph{Planck}}} resolution. This method tries to reproduce the non-gaussian behaviours of the interstellar emission by increasing its fluctuation level as a function of the local brightness \citep{LEA08}.

The thermal Sunyaev-Zel’dovich effect produced by galaxy clusters is also considered. This effect is generally negligible in PS detection, but it is added for completeness. To simulate it, PLA simulations are also used at each frequency, extracting the patches at the same sky position as for the other components (\citealt{Del02}; \citealt{Del13}). Moreover, instrumental noise is added to the simulations by considering white noise using the \textit{\emph{\emph{Planck}}} values: 0.55, 0.78, and 2.56 $\mu K_{CMB}$ deg for 143, 217, and 353 GHz, respectively \citep{PLA18_IV}.

In our work, we considered all the contributions listed above as background ones, except for the PS. Overall, the background at 143 GHz is mainly the emission from the CMB, and it decreases while increasing the frequency. On the other hand, the contamination due to our Galaxy and proto-spheroidal galaxies increases at higher frequencies. 

\begin{figure*}[ht]
\centering
\includegraphics[width=17cm, height=3.5cm]{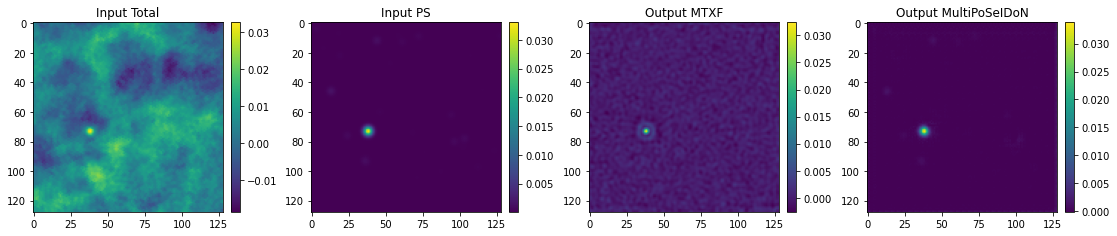}
\includegraphics[width=17cm, height=3.5cm]{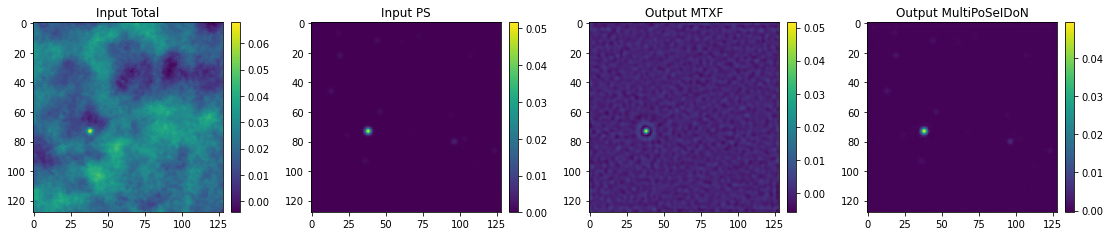}
\includegraphics[width=17cm, height=3.5cm]{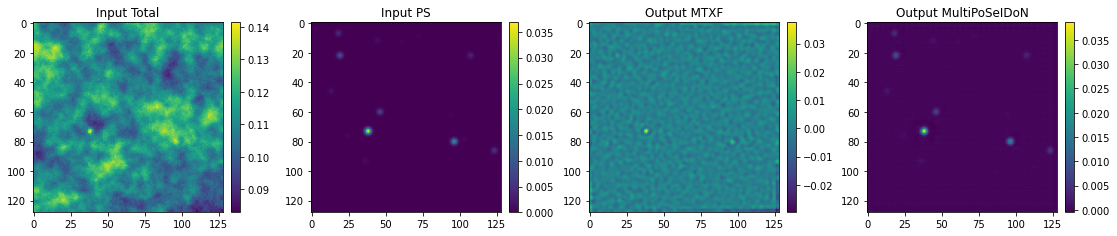}
\caption{From left to right, one simulation for the spectral case with the total and PS-only input validation maps, and the MTXFs and MultiPoSeIDoN PS outputs at 143, 217, and 353 GHz from top to bottom at $b > 30^{\circ}$ Galactic latitudes. The flux density values (in Jy) for each panel are shown in the colour bars.}
\label{Fig 1.}
\end{figure*}

Examples of random simulated patches at the same location in the sky are shown in the first two columns of Figure~\ref{Fig 1.} for 143, 217, and 353 GHz (top, middle, and bottom panels, respectively) at $b > 30^{\circ}$ Galactic latitudes. The first column shows the background with the PS emission (i.e. the total images). The second column represents the PS-only map.

\section{Methodology}
\label{sec:methodology}
\subsection{MultiPoSeIDoN}
\label{sec:FCNN}

Mitchell and Goodfellow 
stated that a computer program is said to learn from experience with respect to a class of tasks and performance measures, if its performance improves with experience (\citealt{Mitchell97}, \citealt{GOO16}). The method by which a computer program learns is called machine learning. Neural networks are machine learning models inspired by the human brain with the goal of learning non-linear behaviours from the data. They are formed by connected layers of neurons, which are their basic computing units. Neurons have weights that adjust their value through a cost function on each step of training. 

When the data have a known grid-like topology, the models used are called convolutional neural networks (CNN, \citealt{LeC89}, \citeyear{LEC15}). In this case, the weights correspond to kernel values, which are tensor-shaped arrays that model the connections between neurons. Every CNN is formed by convolutional blocks. Each of them consist of a layer that performs convolutions in parallel, followed by a set of linear activations, and by a pooling function, which aggregates information by grouping neighbouring pixels generally using their maximum or average values.

The fully-convolutional neural networks (FCN, \citealt{LON15}), which aim to classify each pixel instead of the whole image to perform object segmentation, represent an evolution of this model. They make both learning and inference on the whole image at once through extracting the most relevant characteristics of the image by using convolutional blocks while making a prediction at each pixel by using deconvolutional blocks.

FCN, as any other neural network model, perform optimisation procedures to obtain parameters from data through minimising a loss function. To drive the loss function to a minimum, a gradient-based optimiser is used \citep{Cauchy1847}. When the loss function is adjusted, an algorithm called backpropagation \citep{Rum86} flows the information backward through the network, changing the weights of the neurons. At the same time, the optimiser learns by obtaining an unbiased estimate of the gradient on a small set of samples called minibatch \citep{Kiefer1952}.

FCN are well-suited models for PS detection, as was concluded in \citet{BON21}, since they make inferences through their convolutional blocks to detect PS in noisy maps, while they make predictions through their deconvolutional blocks to produce cleaned PS maps. With these output images, the catalogues can be created and studied statistically.

MultiPoSeIDoN is the FCN developed in this work to detect PS in noisy, multi-frequency background maps. It is a U-Net-based neural network \citep{RON15} with two different approaches: flat and spectral MultiPoSeIDoNs.
The first one is trained with a set of 50 000 simulations of background and PS at 143, 217, and 353 GHz as inputs and a set of 50 000 simulations of PS-only at 217 GHz as labels.
The second one is trained with a set of 50 000 simulations of background and PS and a set of 50 000 simulations of PS-only. In this case, both sets of simulations are at 143, 217, and 353 GHz as inputs and labels, respectively. Therefore, 200 000 and 300 000 images in total are used to train flat and spectral MultiPoSeIDoNs, respectively.
Both training procedures are performed during 500 epochs (when an entire dataset is passed forward and backward through the neural network only once) by using a mean-squared error loss function (MSE, \citealt{has01}) and the adaptive gradient algorithm \citep[AdaGrad,][]{Duchi2011} to perform the learning with a rate of 0.05 on each minibatch of 32 samples. These hyperparameters were selected through a grid search. 

\begin{figure*}[ht]
\centering
\includegraphics[width=17cm]{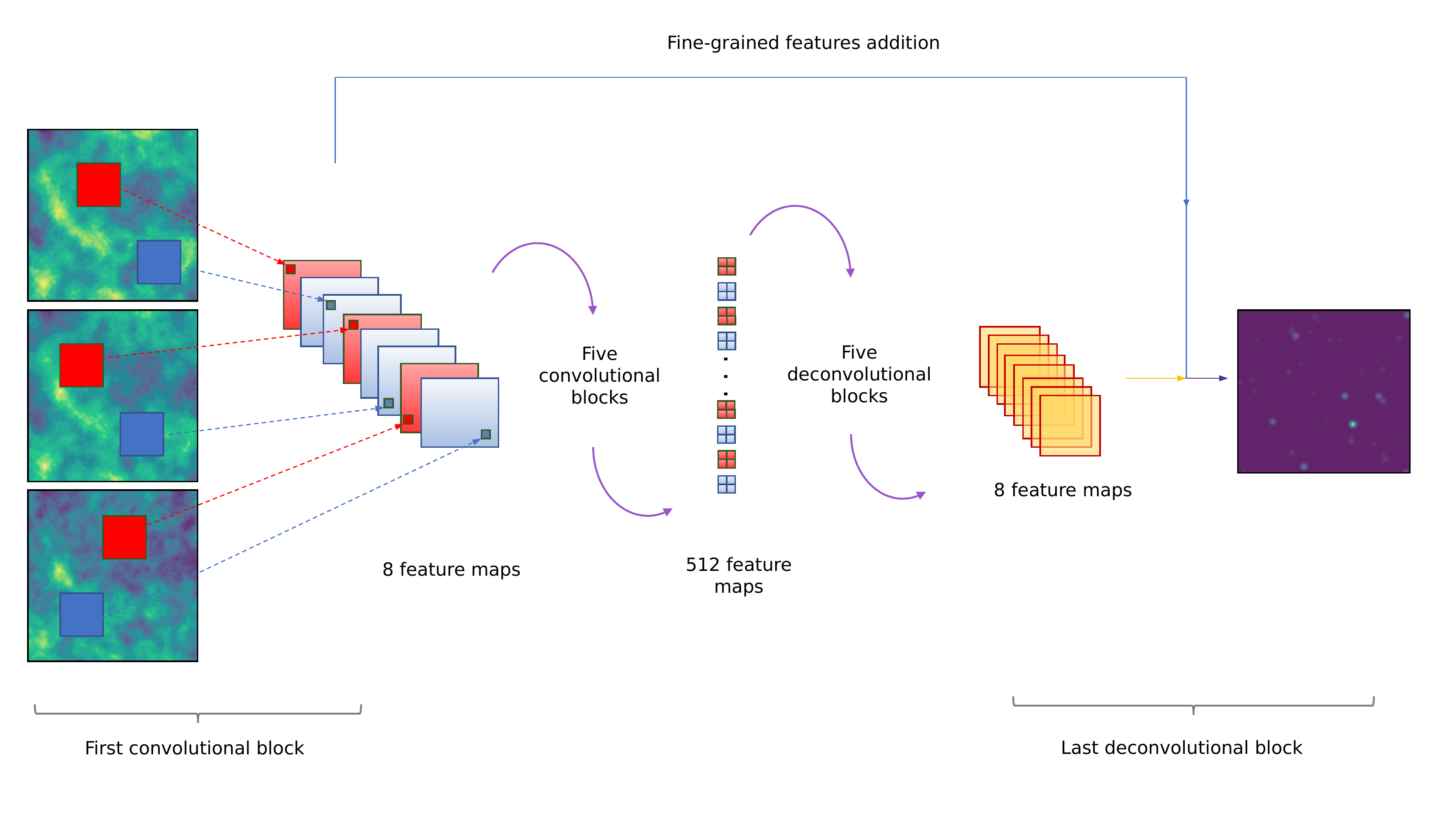}
\includegraphics[width=17cm]{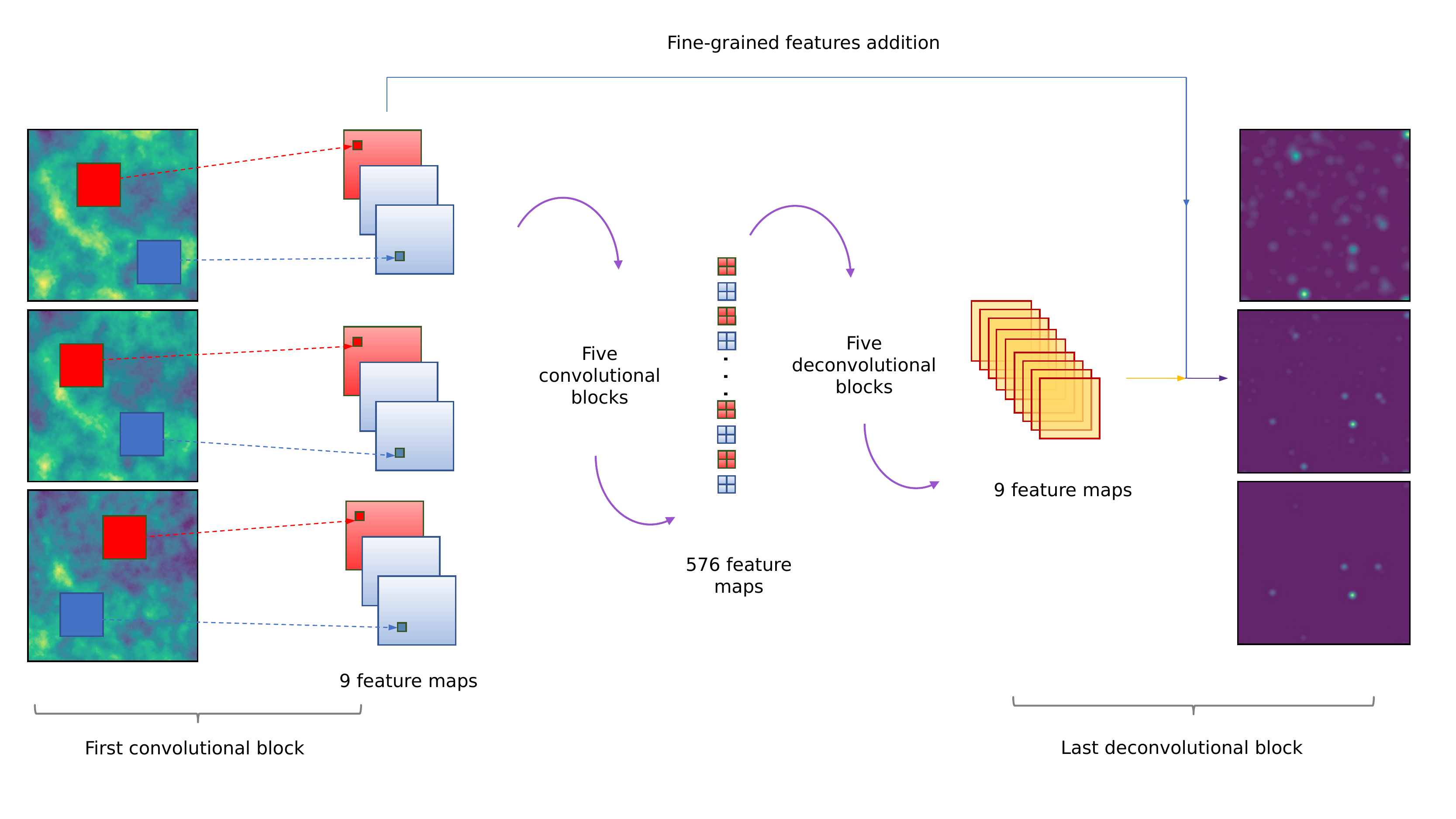}
\caption{Architecture of flat (top panel) and spectral (bottom panel) MultiPoSeIDoNs. The first one has a convolutional block, which produces eight feature maps. After that, the space dimensionality increases to 512 feature maps through five more convolutional blocks. The second one produces 9 and 576 feature maps in its first and last convolutional blocks, respectively. These layers are connected to deconvolutional ones, which decreases the space dimensionality to eight and nine feature maps in the last deconvolutional block for flat and spectral MultiPoSeIDoNs, respectively. Fine-grained features are added from each convolution to its corresponding deconvolution in both neural networks.}
\label{Fig 2.}
\end{figure*}

The architectures for both flat and spectral MultiPoSeIDoNs are detailed as follows (top and bottom pannel in Figure~\ref{Fig 2.} respectively).
First, the neural networks have a set of convolutional blocks: flat MultiPoSeIDoN have six convolutional blocks containing both convolutional and pooling layers with 8, 2, 4, 2, 2, and 2 kernels of sizes of 9, 9, 7, 7, 5, and 3, respectively. Its feature maps are 8, 16, 64, 128, 256, and 512, respectively. On the other hand, spectral MultiPoSeIDoN have the same number of layers, kernels, and kernel sizes, but, since we are using three label images instead of one, as for the flat case, we need to use threefold feature maps. In this case, they are 9, 18, 72, 144, 288, and 576. Both networks have a sub-sampling factor of 2. The padding type 'Same' (an additional layer that adds 'space' around the input data or the  feature map, which helps to deal with possible loss in width and/or height dimensions in the feature maps after having applied the filters) is added in all the layers and the activation function is a leaky ReLU \citep{NAI10}.

After that, the neural networks have a set of deconvolutional blocks: flat MultiPoSeIDoN convolutional blocks are connected to six inverse-convolutional (also called deconvolutional) plus pooling layers with 2, 2, 2, 4, 2, and 8 kernels of sizes of 3, 5, 7, 7, 9, and 9, respectively. Its feature maps are 256, 128, 64, 16, 8, and 1. On the other hand, the spectral MultiPoSeIDoN deconvolutional blocks are formed by the same number of layers, kernels, and kernel sizes, but its feature maps are 288, 144, 72, 18, 9, and 3, respectively. The sub-sampling factor is 2. The padding type Same is added in all the layers, and the activation function is a leaky ReLU. The feature maps resulting from the five last convolutions are added as fine-grained features to the results of the five first deconvolutions.

Both flat and spectral MultiPoSeIDoNs learn through their convolutional blocks that a PS is located at a given position in the background using the position and flux density information provided by the PS-only image, while their deconvolutional blocks perform a PS segmentation from the total input maps, resulting in a PS-only output image. 

Both versions of MultiPoSeIDoN are trained with a large number of simulations to prevent underfitting, and they are tested during training using 5 000 simulations isolated from the training dataset to prevent overfitting. However, since the objective is to predict a numerical flux density of the same type of object (i.e. a point in a map), overfitting is not a problem because the main goal is to deal with background in order to decrease the number of spurious sources (i.e. false positives) instead of detecting different objects in an image. However, learning curves \textbf{of} training and test errors have been used to prevent overfitting during training \citep{GOO16}. An example of a spectral MultiPoSeIDoN output patch (at 143, 217, and 353 GHz, from top to bottom) is shown in the last column of Figure~\ref{Fig 1.}. Similar output images are obtained for the flat MultiPoSeIDoN case.

\subsection{Matched matrix filters}

The matched matrix filters (MTXFs; \citealt{Herranz08}, \citealt{Herranz09}) was introduced
as a method to detect multi-frequency extragalactic PS using their distinctive spatial behaviour, multi-wavelength information, and the fact that their frequency dependence was not known a priori. 

Let us consider a set of $N$ two-dimensional images (channels) formed by PS and other foreground components mainly called generalised noise. The data model is defined by
\begin{equation}
    D_{k} \hspace{1pt} (\mathbf{x}) \hspace{2pt} = \hspace{2pt} s_{k} \hspace{1pt} (\mathbf{x}) \hspace{2pt} + \hspace{2pt} n_{k} \hspace{1pt} (\mathbf{x}),
\label{eq:data_model}
\end{equation}
where $s_{k} \hspace{1pt} (\mathbf{x})$ is the term involving the PS and $n_{k} \hspace{1pt} (\mathbf{x})$ is the one for the generalised noise. The subscript $k = 1, ..., N$ is the index of each image. The term for the PS is defined as
\begin{equation}
    s_{k} \hspace{1pt} (\mathbf{x}) \hspace{2pt} = \hspace{2pt} A_{k} \hspace{1pt} \tau_{k} \hspace{1pt} (\mathbf{x})
\label{eq:point_sources_term}
\end{equation}
where $A_{k}$ is the unknown amplitude of the source in the \textit{k}th channel and $\tau_{k} \hspace{1pt} (\mathbf{x})$ is the known spatial profile of the source. On the other hand, for the generalised noise, the MTXFs assume a zero mean, i.e. $\langle n_{k} \hspace{1pt} (\mathbf{x}) \rangle \hspace{2pt} = \hspace{2pt} 0,$ which can be characterised by its cross-power spectrum:
\begin{equation}
    \langle n_{k} \hspace{1pt} (\mathbf{q}) \hspace{1pt} n_{l}^{*} \hspace{1pt} (\mathbf{q}^{\prime}) \rangle \hspace{2pt} = \hspace{2pt} P_{k l} \hspace{1pt} (\mathbf{q}) \hspace{1pt} \delta^{2} \hspace{1pt} (\mathbf{q} \hspace{1pt} - \hspace{1pt} \mathbf{q}^{\prime}),
\label{eq:cross_power_spectrum}
,\end{equation} 
where $\mathbf{P} \hspace{1pt} = \hspace{1pt} (P_{k l})$ is the cross-power spectrum matrix and '*' denotes complex conjugation.

For accurate photometry in each channel, one has to produce $N$ input maps and other $N$ output maps. For a multi-wavelength approach, one can define a set of $N \times N$ filters $\Psi_{k l}$ , which allows the input channels to help in the elaboration of the output maps. Then, the filtered images for a set of data $D_{l}$ can be defined as
\begin{equation}
\begin{split}
    \omega_{k} (\pmb{x}) \hspace{3pt} & = \hspace{3pt} \sum_{l} \int d \hspace{1pt} \pmb{x^{\prime}} \hspace{1pt} \Psi_{k l} \hspace{1pt} (\pmb{x} - \pmb{x^{\prime}}) \hspace{1pt} D_{l} \hspace{1pt} (\pmb{x^{\prime}}) \\
    & = \hspace{3pt} \sum_{l} \int d \hspace{1pt} \pmb{q} \hspace{1pt} e^{-i \hspace{1pt} \pmb{q} \hspace{1pt} \pmb{x}} \hspace{1pt} \Psi_{k l} \hspace{1pt} (\pmb{q}) \hspace{1pt} D_{l} \hspace{1pt} (\pmb{q}),
\end{split}
\label{eq:sum_set_linear_filters}
\end{equation}
where $\omega_{k}$ is the sum of a set of linear filters of the data $D_{l}$. The root mean square (rms) of the filtered data is then estimated using the square root of the variance: 
\begin{equation}
\sigma_{\omega_{k}}^{2} \hspace{3pt} = \hspace{3pt} \sum_{l} \hspace{1pt} \sum_{m} \int d \hspace{1pt} \pmb{q} \hspace{1pt} \Psi_{k l} \hspace{1pt} (\pmb{q}) \hspace{1pt} \Psi_{k m}^{*} \hspace{1pt} P_{l m} \hspace{1pt} (\pmb{q}).
\label{eq:rms_filtered_data}
\end{equation}
The set of filters $\mathbf{\Psi}$ that minimise the variance $\sigma_{\omega_{k}}$ for all $k$ channels, and for the same amplitudes $A_{k}$ of the sources, as was derived in \citet{Herranz08}, can be obtained by matrix multiplication:
\begin{equation}
    \mathbf{\Psi}^{*} \hspace{3pt} = \hspace{3pt} \mathbf{F} \hspace{1pt} \mathbf{P}^{-1},
\label{eq:filters_minimize_variance}
\end{equation}
where
\begin{equation}
\begin{split}
    \mathbf{\Psi} \hspace{2pt} & = \hspace{2pt} (\Psi_{k l}), \hspace{2pt} \mathbf{F} \hspace{2pt} = \hspace{2pt} (F_{k l}), \hspace{2pt} \mathbf{P} \hspace{2pt} = \hspace{2pt} (P_{k l}), \\
    \pmb{\lambda} \hspace{2pt} & = \hspace{2pt} (\lambda_{k l}), \hspace{2pt} \mathbf{H} \hspace{2pt} = \hspace{2pt} (H_{k l}), 
\label{eq:filter_matrices}
\end{split}
\end{equation}
are $N \times N$ matrices for any $\pmb{q,}$ and
\begin{equation}
\begin{split}
    & F_{k l} \hspace{2pt} = \hspace{2pt} \lambda_{k l} \hspace{1pt} \tau_{l}, \\
    & \pmb{\lambda} \hspace{2pt} = \hspace{2pt} \mathbf{H}^{-1}, \\ & H_{k l} \hspace{2pt} = \hspace{2pt} \int d \hspace{1pt} \pmb{q} \hspace{1pt} \tau_{k} \hspace{1pt} (\pmb{q}) \hspace{1pt} P_{k l}^{-1} \hspace{1pt} \tau_{l} \hspace{1pt} (\pmb{q}).
\label{eq:filter_matrices_2}
\end{split}
\end{equation}
When the filters \eqref{eq:filters_minimize_variance} are applied to a set of $N$ images, the filtered ones with values $[\omega_{1},...,\omega_{N}]$ have the form
\begin{equation}
\omega_{k} \hspace{3pt} = \hspace{3pt} \sum_{l} \int d \hspace{1pt} \pmb{q} \hspace{1pt} \Psi_{k l} \hspace{1pt} (\pmb{q}) \hspace{1pt} A_{l} \hspace{1pt} \tau_{l} \hspace{1pt} (\pmb{q}) \hspace{3pt} = \hspace{3pt} A_{k}.
\label{eq:filtered_images}
\end{equation}
To avoid the fact that the flux density of a PS in the \textit{l}th channel could leak into the \textit{k}th filtered image, the filters satisfy an orthonormality condition:
\begin{equation}
    \int d \hspace{1pt} \pmb{q} \hspace{1pt} \Psi_{k l} \hspace{1pt} (\pmb{q}) \hspace{1pt} \tau_{l} \hspace{1pt} (\pmb{q}) \hspace{2pt} = \hspace{2pt} \delta_{k l}.
\label{eq:orthonormality_condition_filters}
\end{equation}
Therefore, the filters are \textit{\emph{unbiased}} estimators of the flux density of the PS for all $N$ channels. Moreover, to avoid the possibility of statistical correlations between the generalised noise in the \textit{l}th and \textit{k}th channels, the $H_{k l}$ terms in \eqref{eq:filter_matrices_2} minimise the filtered rms $\sigma_{\omega_{k}}$ to lower values than the ones achieved by single-frequency matched filters (\citealt{Herranz08}, \citealt{Herranz09}). 

Where the noise is uncorrelated through channels, the matrix of filters are diagonal matrices with elements:
\begin{equation}
    \Psi_{k l}^{*} \hspace{1pt} (\pmb{q}) \hspace{2pt} = \hspace{2pt} \delta_{k l} \frac{\tau_{k} \hspace{1pt} (\pmb{q}) \hspace{1pt} / \hspace{1pt} P_{k} \hspace{1pt} (\pmb{q}) \hspace{1pt}}{\int d \hspace{1pt} (\pmb{q}) \hspace{1pt} \tau_{k}^{2} \hspace{1pt} (\pmb{q}) \hspace{1pt} / \hspace{1pt} P_{k} \hspace{1pt} (\pmb{q})},
\label{eq:diagonal_elements_filters}
\end{equation}
that is, the non-zero elements of the matrix filters are the complex conjugates of the matched filters of each channel. When the PS have symmetric profiles or statistically homogeneous and isotropic noise, the MTXFs show the same performance as the single-frequency matched filter, which is formed by real values. Since for real foreground-contaminated data, the generalised noise is neither homogeneous nor isotropic and is further correlated between channels, MTXFs are one of the best-suited methods to detecting multi-frequency PS.

With all these assumptions, one can conclude that MTXFs detect PS by removing the generalised noise by filtering in the scale domain of the sources and by cleaning out large-scale structures localised in neighbouring channels. An example of MTXF's output patch (at 143, 217, and 353 GHz, from top to bottom) is shown in the third column of Figure~\ref{Fig 1.}. 

\section{Results}
\label{sec:results}

Once we obtained the outputs of MultiPoSeIDoN and the MTXFs on the same validation simulations (a set of 5 000 simulations isolated from train and testing datasets), we created the input and output (with the detections) catalogues by searching the local maxima of the sources, after introducing a threshold in flux density of 60 mJy for the three models, and also a safer 4$\sigma$ limit for the MTXFs, as the one used in \textit{\emph{\emph{Planck}}} catalogues. We also considered a detection when sources are separated by a two-pixel minimum distance.
Border effects are one of the main problems of filtering techniques such as the MTXFs, because the detection algorithm could consider some artefacts near the patch border as spurious PS detections. As was concluded in \citet{BON21}, neural networks does not have to deal with border effect issues, so the whole patch can be included in the analysis. However, for the MTXFs, some pixels in the patch border must be avoided. In our case, we removed five pixels both in width and height.
To asses our results, we performed a statistical analysis of the catalogues resulting from the outputs of both flat and spectral MultiPoSeIDoNs, as well as for the MTXFs.
The statistical quantities analysed are the completeness, the percentage of spurious sources, and the flux density comparison between the input and the recovered values (\citealt{LOP07}; \citealt{ERCSC}, \citeyear{PCCS}, \citeyear{PCCS2}, \citeyear{PLA18_I}; \citealt{Hop15}). 
Completeness is the ratio between the number of recovered true sources and the total number of input sources over a
given flux limit. It is defined by the relation
\begin{equation}
    C \hspace{1pt} (> S_{0}) \hspace{2pt} = \hspace{2pt} \frac{N_{true \hspace{1pt} detected}  \hspace{1pt} (> S_{0})}{N_{input} \hspace{1pt} (> S_{0})},
\label{eq:completeness}
\end{equation}
where $S_{0}$ is the input flux density, $N_{true \hspace{1pt} detected} \hspace{1pt} (> S_{0})$ is the number of true detected sources by each method, and $N_{input} \hspace{1pt} (> S_{0})$ is the number of sources in the PS-only input catalogue. 
The detected sources that do not have a counterpart in the input catalogue (i.e. false positives) are called spurious sources. 
Their number can be estimated using
\begin{equation}
\begin{split}
    R \hspace{1pt} (S_{0}) \hspace{2pt} & = \hspace{2pt} \frac{N_{spurious} \hspace{1pt} (> S_{0})}{N_{input} \hspace{1pt} (> S_{0})} \hspace{2pt} \\
    & = \hspace{2pt} \frac{N_{output} \hspace{1pt} (> S_{0}) \hspace{1pt} - \hspace{1pt} N_{true \hspace{1pt} detected} \hspace{1pt} (> S_{0})}{N_{input} \hspace{1pt} (> S_{0})},
\end{split}
\label{eq:spurious_sources}
\end{equation}
where $S_{0}$ is the input flux density, $N_{output} \hspace{1pt} (> S_{0})$ is the number of detected sources after having used each method, $N_{true \hspace{1pt} detected} \hspace{1pt} (> S_{0})$ is the number of detected sources which are in both the input and detection catalogues (i.e. true detections), and $N_{input} \hspace{1pt} (> S_{0})$ is the number of sources in the PS-only input catalogue. 

\begin{figure*}[ht]
\centering
\subfigure{\includegraphics[width=8.0cm]{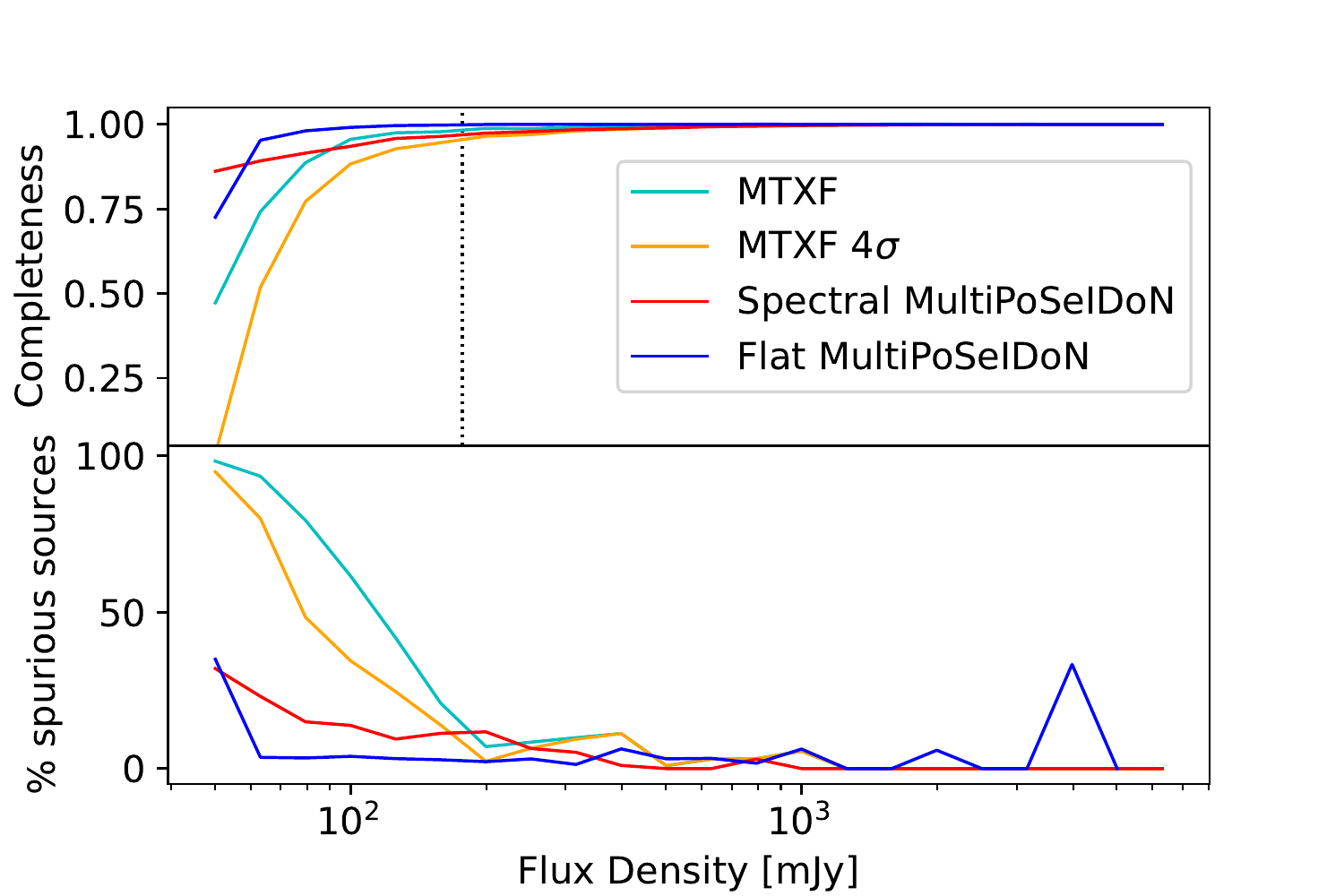}}
\qquad
\subfigure{\includegraphics[width=8.0cm]{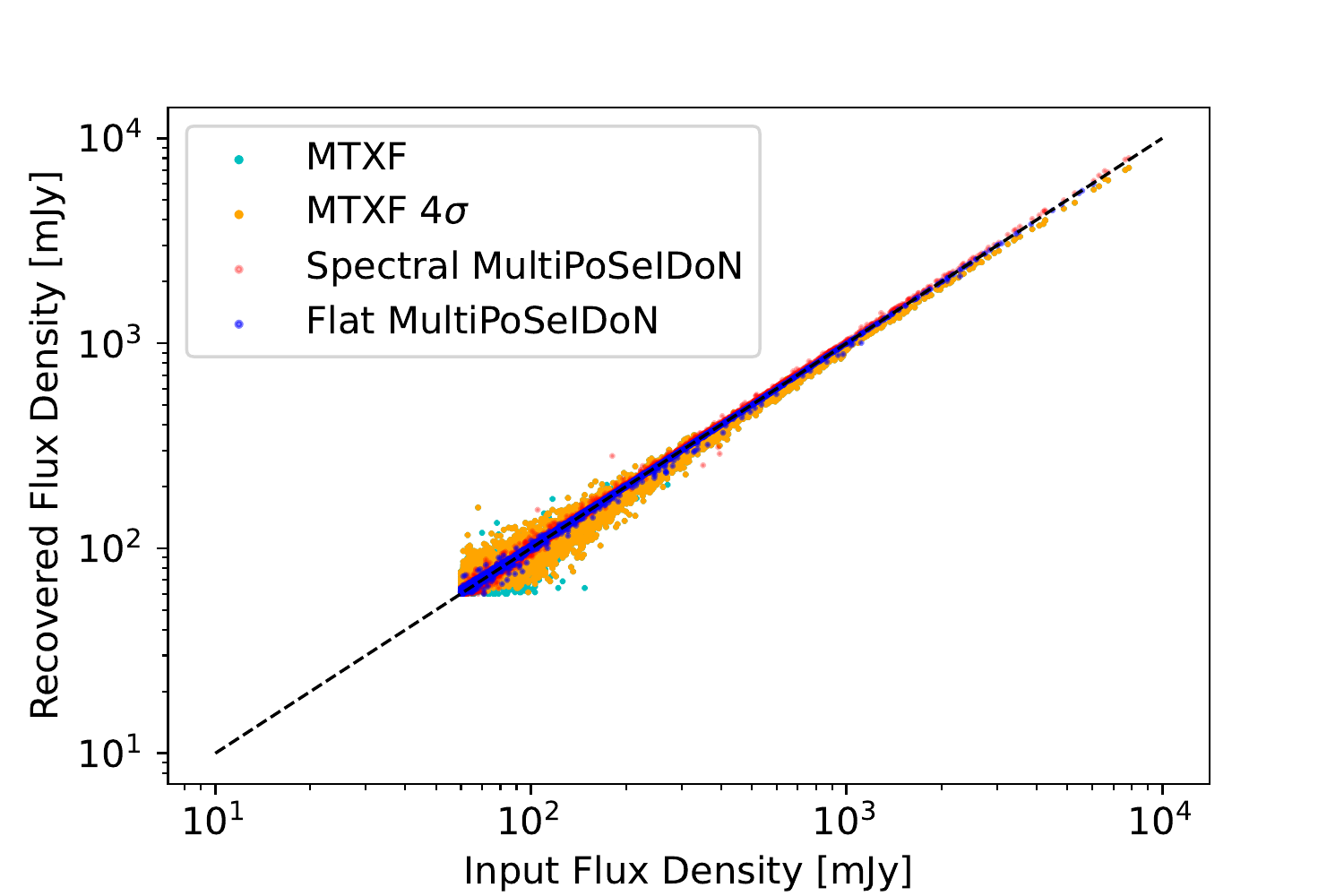}}

\subfigure{\includegraphics[width=8.0cm]{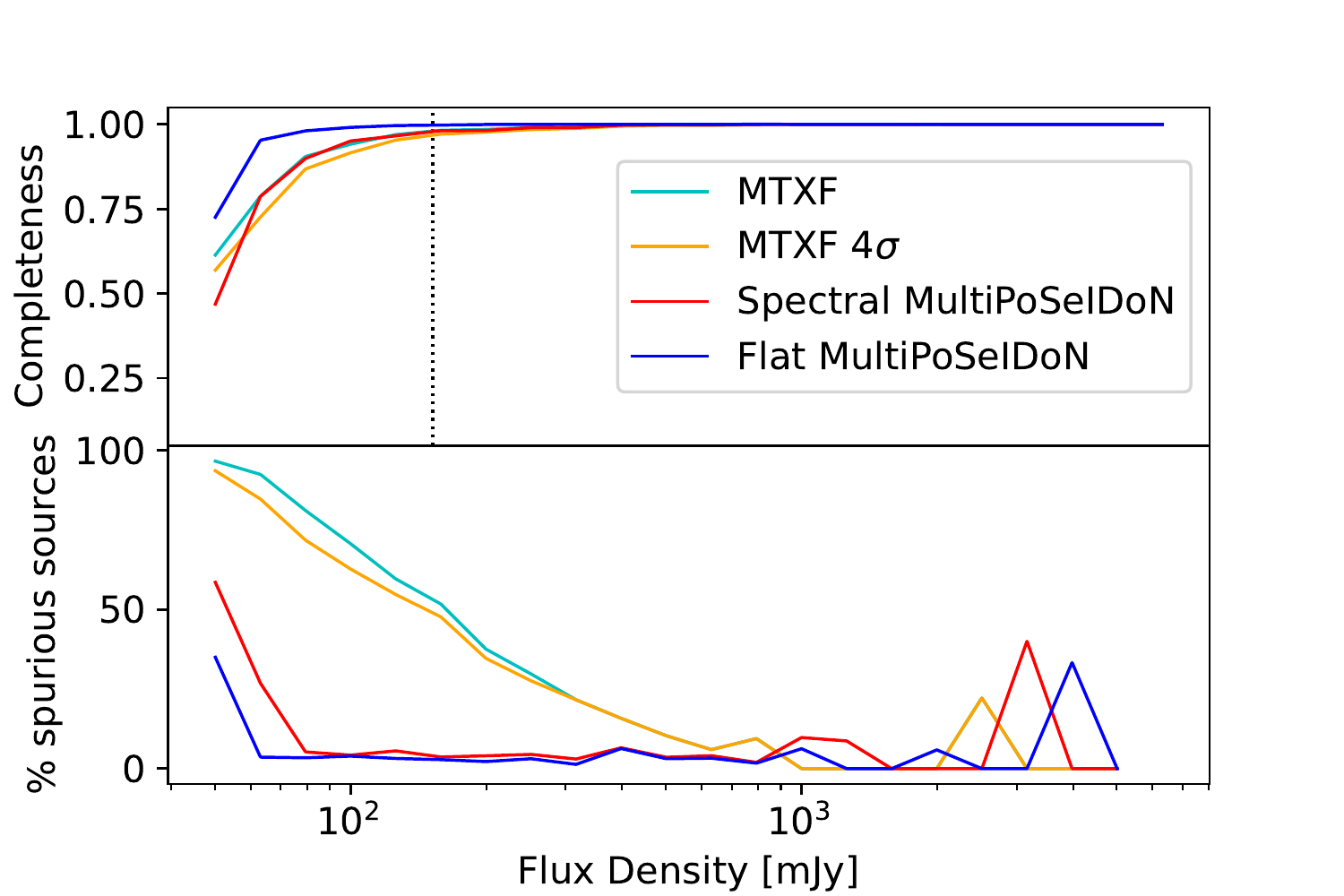}}
\qquad
\subfigure{\includegraphics[width=8.0cm]{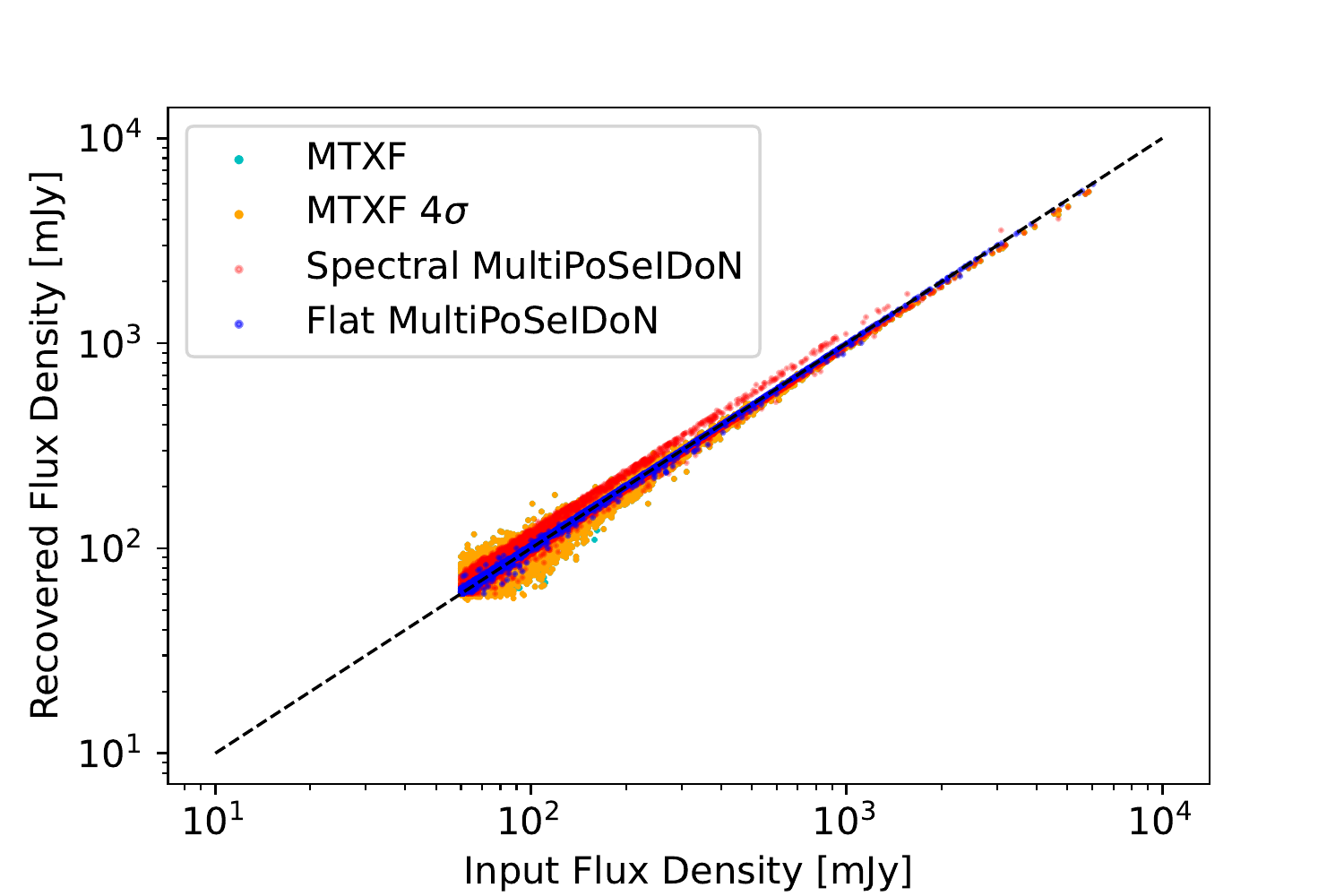}}

\subfigure{\includegraphics[width=8.0cm]{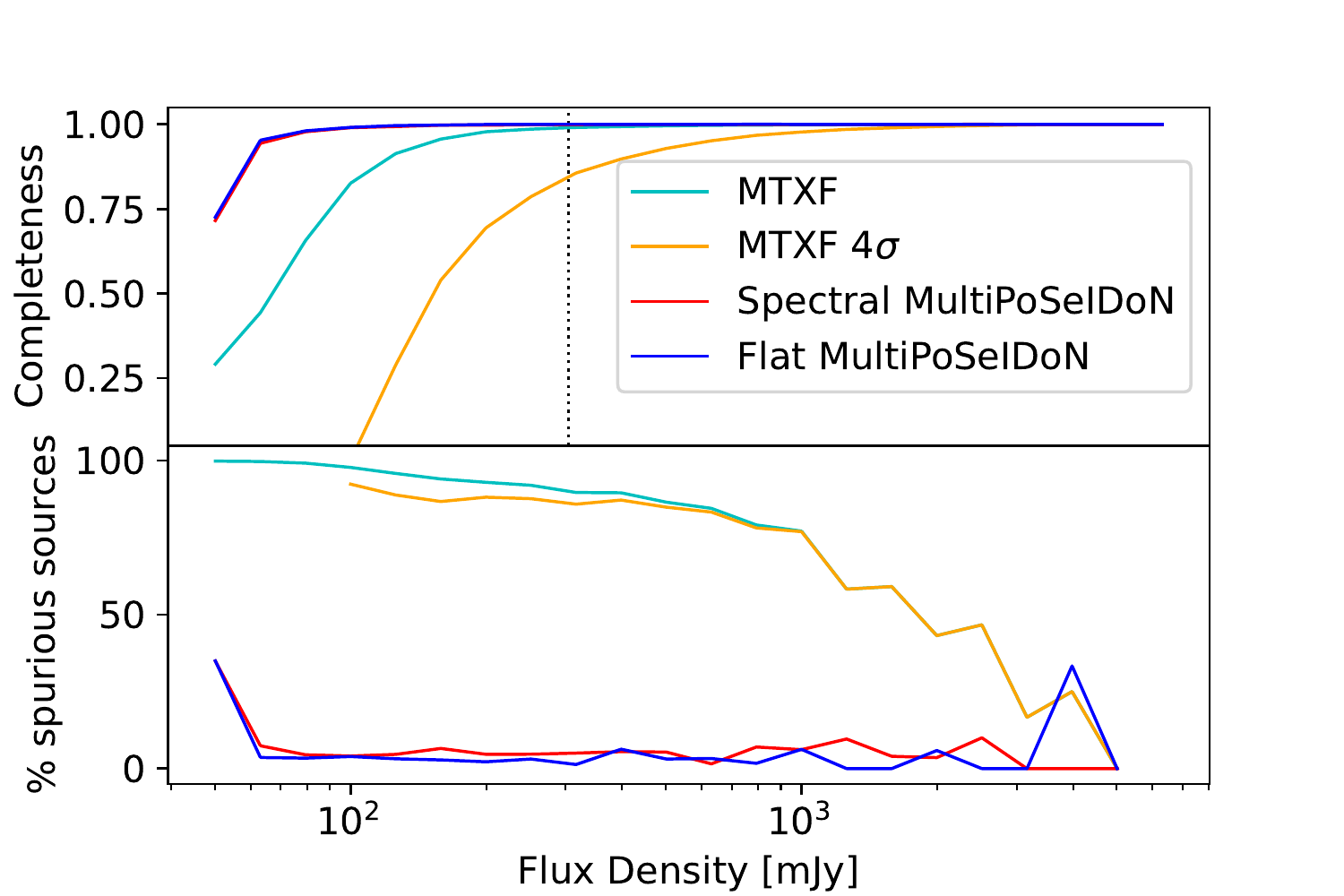}}
\qquad
\subfigure{\includegraphics[width=8.0cm]{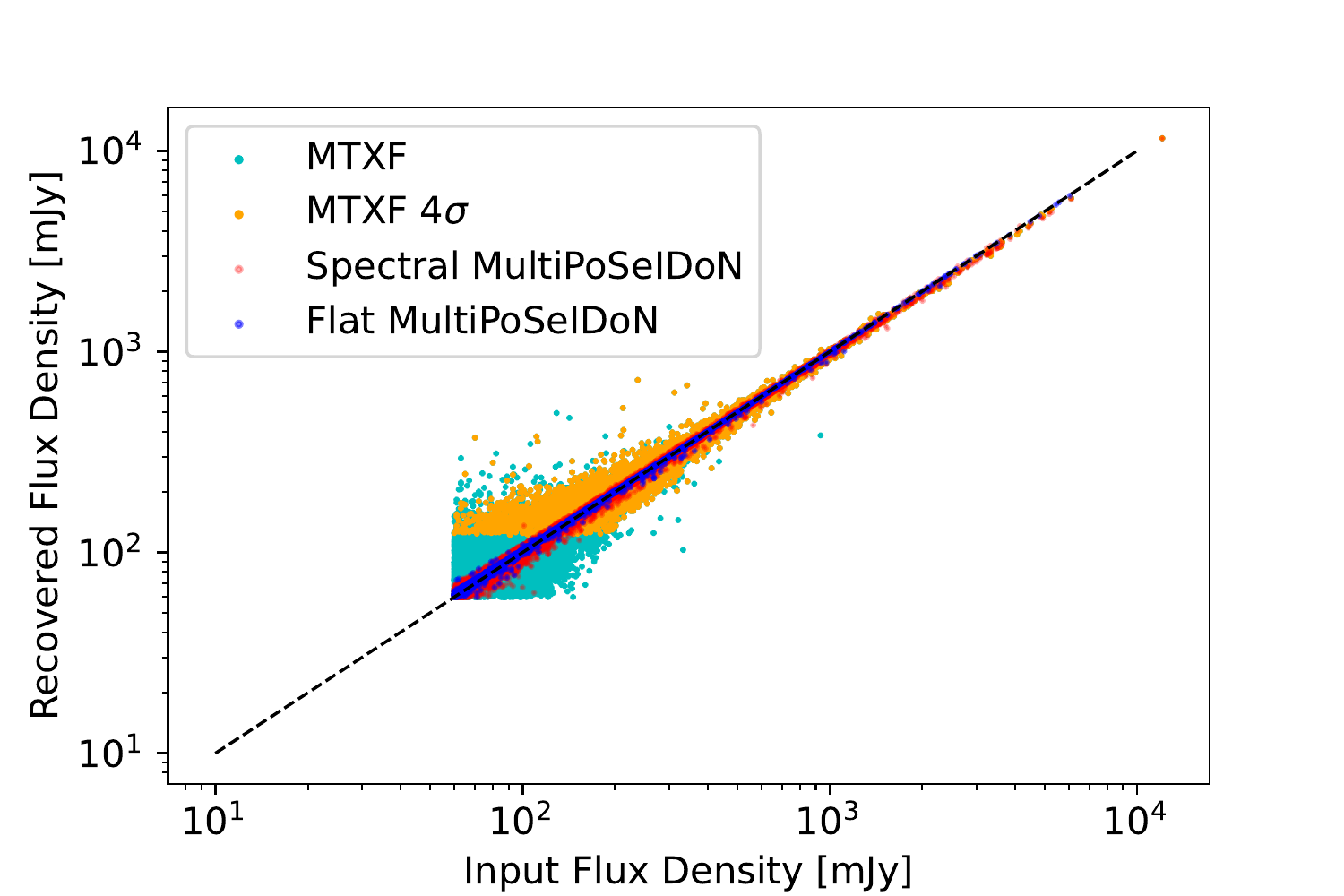}}


\caption{Validation results of completeness (left column, top sub-panel), percentage of spurious sources (left column, bottom sub-panel), and flux density comparison (right column) between the input sources and the recovered ones by flat (blue series) and spectral (red series) MultiPoSeIDoNs, the MTXFs with a detection limit of 60 mJy (cyan series), and with a 4$\sigma$ detection limit (orange series). The frequencies are 143 (top panels), 217 (middle panels), and 353 GHz (bottom panels) for a $30^{\circ}$ Galactic cut. The dotted grey vertical line is the 90\% completeness flux density limit (177, 152, and 304 mJy for 143, 217, and 353 GHz, respectively) for the second \textit{\emph{\emph{Planck}}} catalogue of compact sources \citep{PCCS2}.}
\label{Fig 3.}
\end{figure*}

\subsection{Completeness and spurious sources}

The completeness (top sub-panel) and the percentage of spurious sources (bottom sub-panel) with respect to the input ones are shown on the left column of Figure~\ref{Fig 3.}.
For the MTXFs (cyan solid line), we obtain the expected results: using a detection limit of 60 mJy in flux density, the MTXFs provide good completeness results for the three \textit{\emph{\emph{Planck}}} channels, reaching the 90\% completeness level at 84, 79, and 123 mJy for 143, 217, and 353 GHz, respectively. We also used a 4$\sigma$ limit (orange solid line) to reduce the number of spurious detections, reaching the 90\% completeness level at 113, 92, and 398 GHz.

The percentage values of spurious sources are sensitive to detection limits in flux density; more than 20\% of the total detected sources are spurious bellow 180 mJy at 143, and this behaviour increases with frequency, reaching the same percentage value below $\sim$400 and 1200 mJy for 217 and 353 GHz, respectively. The application of a 4$\sigma$ detection limit for the flux density does not improve these results much.

Flat (blue solid line) and spectral (red solid line) MultiPoSeIDoNs have a relatively similar completeness results at 143 and 217 GHz, reaching the 90\% completeness level at 58 mJy for a flat MultiPoSeIDoN and 79 and 71 mJy for a spectral MultiPoSeIDoN at 143 and 217 GHz, respectively. For 353 GHz, the completeness results of MultiPoSeIDoN are much better than for the MTXFs, reaching the the 90\% completeness level at 60 mJy.
The clear advantage of MultiPoSeIDoN with respect to the MTXFs is shown in the percentage of spurious sources panel: both neural networks reach percentages of spurious sources of 20\% below 100 mJy. Moreover, above the 90\% of completeness level, they detect sources with a percentage of spurious sources between 0 and 10\%. This behaviour is similar to that found in \citet{BON21}, when a single-frequency neural network called PoSeIDoN was compared in completeness and percentage of spurious sources with the Mexican hat wavelet 2 (MHW2 \citealt{GN06}). The spurious PS issue is mainly due to a strongly contaminant background when the emission from our Galaxy and the CIB is higher. Therefore, obtaining lower values of spurious sources implies that MultiPoSeIDoN detects PS with higher accuracy than the MTXFs.

At 217 GHz, there is a peak in both flat and spectral cases, showing a high number of spurious sources above 1 000 mJy up to about 4 000 mJy. These peaks correspond to sources located at the borders of the map. This small issue is due to the catalogue production on 
the MultiPoSeIDoN outputs rather than the MultiPoSeIDoN detection capability.

By comparing the two versions of MultiPoSeIDoN, the flat one performs slightly better than the spectral one based on 
the results of completeness and reliability. This is explained by considering that spectral MultiPoSeIDoN have to deal with the spectral behaviour of PS, challenging their detection, as in a realistic case.

\subsection{Photometry}

In Figure~\ref{Fig 3.} (right column), we compare the recovered and input flux density of the sources.
On one hand, the MTXFs correctly recover most of the flux densities above 100 mJy at 143 and 217, respectively, and above 150 mJy at 353 GHz within a 10\% relative error. However, below those flux density values, the filters perform with a higher relative error, mainly due to very faint PS that are hardly detectable and could be located near a CMB fluctuation or a region with a high contamination from our Galaxy. This effect is called the Eddington bias \citep{EDD13}, which is more visible at 353 GHz, when the contamination from the background is higher.

On the other hand, both flat (blue dots) and spectral 
(red dots) MultiPoSeIDoNs recover the flux density of the sources quite well, especially the brighter ones (flux densities above 100 mJy), the flat version being slightly better than the spectral one, as explained in the previous section. However, the results are still very good even at the fainter flux densities (already at 100 mJy for all the frequencies) with relative errors such as 5\% for the flat case and between 7\% and 10\% for the spectral case. 

At 217 GHz, a double diagonal can be seen above 300 mJy with a relative error of $\pm$5\%. This issue is due to a bad fitting of the fainter IRLT sources by the network. At this frequency, either IRLT and radio galaxies are in the training catalogue, as was explained before. More particularly, in this regime, the IRLT population is generally formed by fainter sources. This bad fitting is due to a not sufficiently accurate assigned spectral index value for the fainter IRLT sources, implying a worse performance recovering the flux density of the sources for the neural network at this frequency and flux density regime. However, MultiPoSeIDoN is removing part of the background that is increasing the flux density of those sources, which is the main objective of the network, and the detection is not affected at all by this issue attending on the results of completeness and spurious detections.

\subsection{Multi-frequency versus single-frequency point source detection with fully convolutional networks}

\begin{figure*}[ht]
\centering


\subfigure{\includegraphics[width=8.0cm]{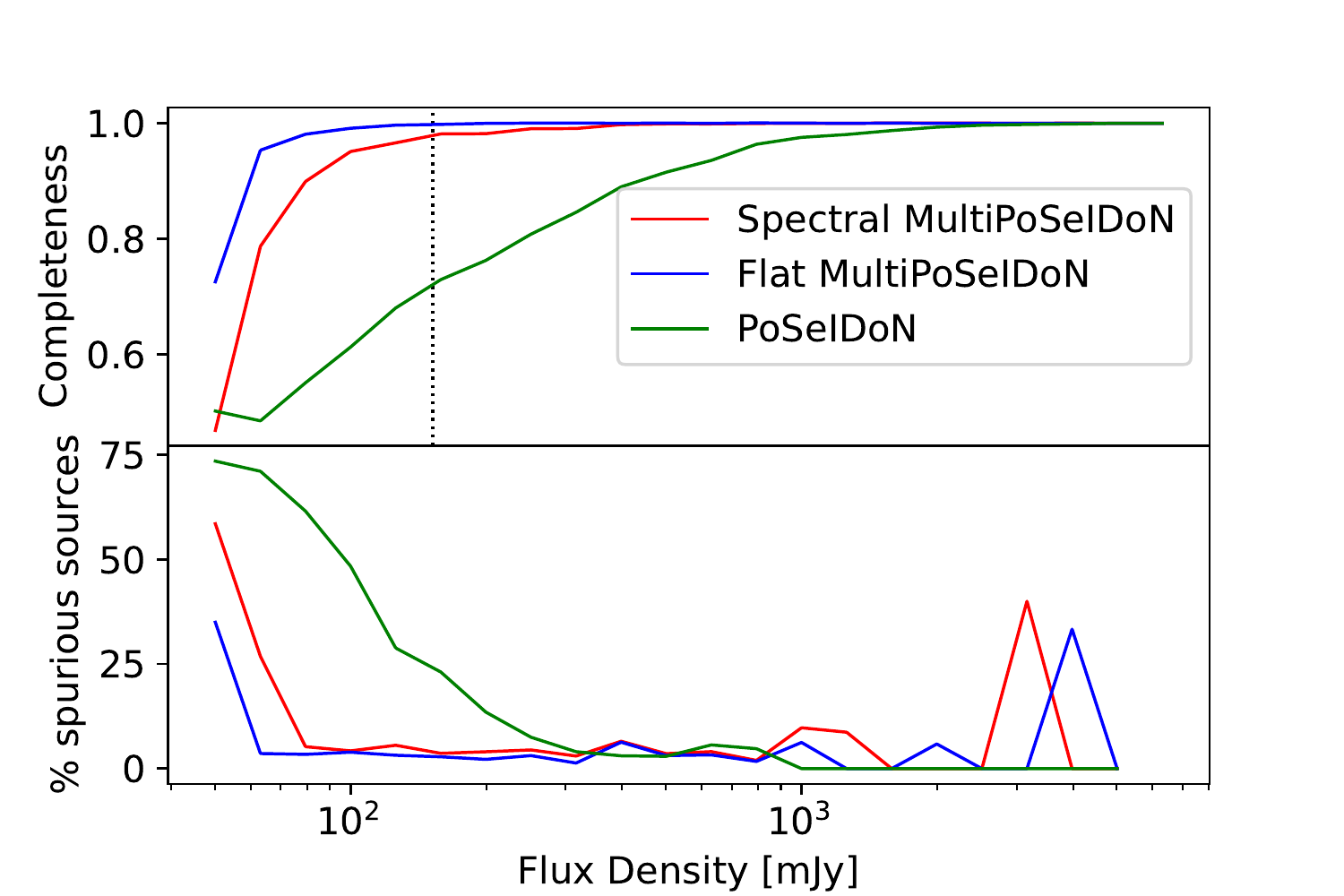}}
\qquad
\subfigure{\includegraphics[width=8.0cm]{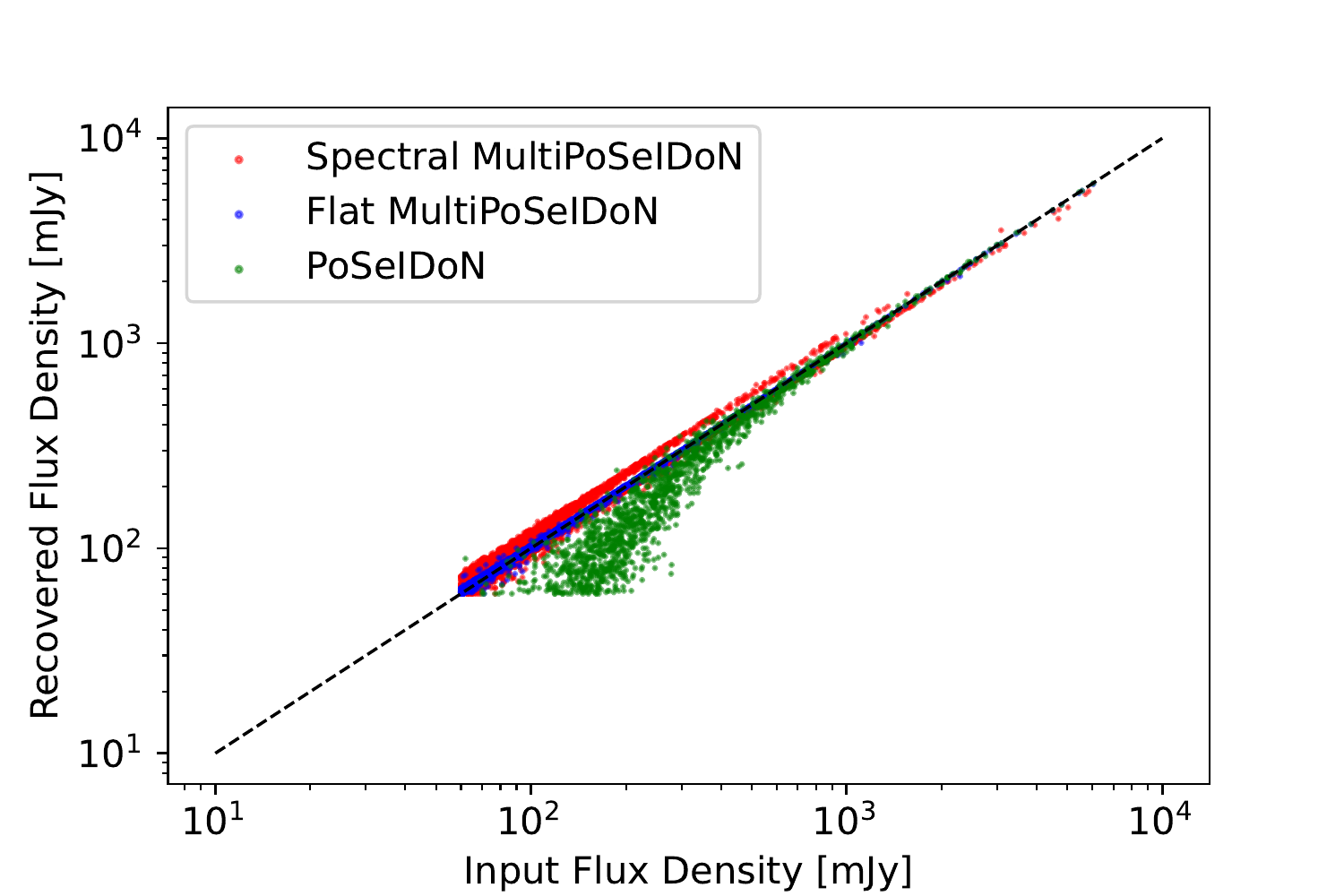}}

\caption{Validation results of completeness (left column, top sub-panel), percentage of spurious sources (left column, bottom sub-panel), and flux density comparison (right column) between the input sources and the recovered ones by Flat (blue series) and Spectral (red series) MultiPoSeIDoNs and PoSeIDoN (green series). The frequency is 217 GHz for a $30^{\circ}$ Galactic cut. The dotted grey vertical line is the 90\% completeness flux density limit (152 mJy) for the second \textit{\emph{\emph{Planck}}} catalogue of compact sources \citep{PCCS2}.}
\label{Fig 4.}
\end{figure*}

PoSeIDoN is a fully convolutional neural network designed to detect PS in single-frequency noisy background maps. Its performance was evaluated in \citet{BON21} by training it at 217 GHz with 50 000 realistic simulated images of PS plus contaminants such as CMB, CIB and thermal dust emissions, the thermal Sunyaev-Zel'dovich effect, and instrumental noise. It learned to detect PS with the help of 50 000 label images formed by PS only (radio and IRLT galaxies). Its performance was compared to the Mexican hat wavelet 2 filter (MWH2, \citealt{GN06}), one of the most popular single-frequency PS detection methods, which was used in the \textit{\emph{\emph{Planck}}} experiment to validate both LFI and HFI channels and to create both PCCS and PCCS2 (\citealt{PCCS}; \citealt{PCCS2}). The results have shown that PoSeIDoN has a similar completeness results to the filter, but with much lower number of spurious detections, not only at the frequency at which it was trained, but also at 143 GHz and 353 GHz. Therefore, they concluded that PoSeIDoN is a more accurate and robust method.

The three statistical quantities explained above are used to compare the performance between PoSeIDoN and MultiPoSeIDoN. Both neural networks are validated with the same 5 000 simulations at 217 GHz. The detection limit of 60 mJy in flux density and the two-pixel minimum distance between the sources to create the detection catalogues are also the same for both models.

As is shown in Figure~\ref{Fig 4.}, the completeness results (left column, top sub-panel) show that both flat and spectral MultiPoSeIDoNs (blue and red solid lines respectively) show better performance than the single-frequency neural network (green solid line), reaching the 90\% completeness level at 58 and 71 mJy, respectively, while PoSeIDoN reaches it at $\sim$300 mJy. On the other hand, the percentage of spurious sources (left column, bottom sub-panel) shows that the detections of MultiPoSeIDoN are more accurate since 20\% of the PoSeIDoN detections are spurious below 150 mJy, while for MultiPoSeIDoN the same percentage of spurious sources is obtained below 100 mJy.

The comparison between the flux density of input sources (from the validation set) and of detected sources is shown at the right column in Figure~\ref{Fig 4.} for the three neural networks. On one hand, PoSeIDoN (green dots) tends to underestimate the flux density of the recovered fainter sources (which was exactly the opposite behaviour to the MHW2, as is shown in \citet{BON21}). On the other hand, both flat and spectral MultiPoSeIDoNs (blue and red dots, respectively) show much better performance in recovering the flux density of the fainter PS, with only a relative error between 5\% and 10\% below 100 mJy, while PoSeIDoN has an up to 80\% relative error below the same flux density limit. For the more intense sources, the three neural networks recover the flux density of the detections quite well.


\section{Conclusions}
\label{sec:conclusions}

In this work, we successfully applied a modern approach for the multi-frequency detection of point sources (PS) based on fully-convolutional neural networks trained with realistic simulations of images of the microwave sky. The patches of the images have contributions from PS (radio and IR late-type galaxies), the emission from the CMB and from the thermal dust of our Galaxy, the contamination due to massive proto-spheroidal galaxies (CIB), the thermal Sunyaev-Zel'dovich effect, and instrumental noise using \textit{\emph{\emph{Planck}}} values. The CMB maps are the ones from the SEVEM method, and the Galaxy and Sunyaev-Zel'dovich contributions are those provided by the PLA. The frequencies were 143, 217, and 353 GHz at $b > 30^{\circ}$ Galactic latitudes.

A total of 50 000 simulations of background plus PS and 50 000 PS-only simulations were used to train two neural networks: flat MultiPoSeIDoN is trained by assuming sources with a flat spectrum (i.e. the sources had the same emission in all the channels), and spectral MultiPoSeIDoN uses simulations considering sources with a spectral behaviour. In that case, the emission from radio galaxies is higher at 143 GHz, whereas at 353 GHz IR late-type galaxies have more impact on the maps. The spectral behaviour is simulated by using the spectral index of each population after fitting their gaussian distributions plotted in PCCS2 \citep{PCCS2}.

After the training, 5 000 simulations (isolated from the training dataset) are used to validate both flat and spectral MultiPoSeIDoNs and our comparison method, the matrix filters (MTXFs). Then, they are compared using their completeness, percentage of spurious sources, and flux density estimation values. Applying a detection limit of 60 mJy, MultiPoSeIDoN performs better than the MTXFs, especially when dealing with spurious sources. It reaches the 90\% completeness level at 58 mJy for the flat case and 79, 71, and 60 mJy for the spectral case at 143, 217, and 353 GHz, respectively. The filter reaches the 90\% completeness level at 84, 79, and 123 mJy. Of these detections, the percentage of spurious ones detected by the neural networks is 20\% below a flux density of 100 mJy, decreasing it to values between 0 and 10\% above that flux density value. For the filter, the percentage of spurious detections is higher than 20\% below 180 mJy at 143 GHz. For the other frequencies, more than 20\% of detections are spurious below $\sim$400 and 1200 mJy. Using the 4$\sigma$ limit does not help to improve these results.

Based on these results, MultiPoSeIDoN has a similar completeness to the MTXFs at 143 and 217 GHz, but it is better at 353 GHz. For all those detections, the ones from the neural network are more reliable, especially for the sources with higher intensity, although we used a safer 4$\sigma$ detection limit. As expected, the performance of the matrix filter is worse at higher frequencies, when the contamination from our Galaxy and the CIB is higher. However, the neural network demonstrates a rather equal performance at these higher frequencies. Based on that behaviour, we can assume that the parameters that model the spectral behaviour of the foreground components are successfully learned by MultiPoSeIDoN. Therefore, we anticipate that the higher contamination from the foregrounds would only weakly affect the performance of the neural network at higher frequencies than 353 GHz, or in regions inside the Galactic mask, where the contamination from our Galaxy has limited the \textit{\emph{\emph{Planck}}} Collaboration when it comes to creating a reliable catalogue of compact sources in those regions. However, those two possibilities are outside the scope of this work.

The results on photometry show that both flat and spectral MultiPoSeIDoNs perform well in recovering the flux density of the sources, especially the brighter ones. On the other hand, the MTXFs are affected by the Eddington Bias at higher flux densities than the neural networks. 

Another advantage of MultiPoSeIDoN with respect to the MTXFs is that we do not need to subtract pixels from the patch borders to avoid the border effect issue of the filtering methods. Therefore, the whole patch was considered in the analysis, allowing MultiPoSeIDoN to detect sources located at the patch border.

The comparison between both neural networks shows that flat MultiPoSeIDoN performs slightly better than the spectral case for the three statistical quantities. However, spectral MultiPoSeIDoN successfully obtains results considering that it is a more general and realistic method. 

MultiPoSeIDoN can be considered as an evolution of its single-frequency relative, PoSeIDoN, because the performance of the multi-frequency approach is better than the one in the single-frequency case, especially regarding the photometry results: MultiPoSeIDoN recovers the flux density of the fainter sources with a much lower relative error (a mean difference of $\sim$70\% with respect the results of PoSeIDoN). There are two main reasons for that: the first one is the increasing of the training information. The flat case was trained with twice the amount of information for PoSeIDoN, and the spectral one had three times the information. The second reason is that MultiPoSeIDoN learned the different correlations between the elements in the simulations due to their spectral behaviours. 

As a future perspective, MultiPoSeIDoN could be trained for other CMB experiments such as the Q-U-I JOint Tenerife experiment (QUIJOTE, \citealt{Rub12}), LiteBird, or PICO. It could be used to study CMB P-polarization modes to compare its performance against a recent Bayesian method, which uses images of realistic simulations with P-polarized compact sources \citep{HER21}. Another interesting study could be to use MultiPoSeIDoN in multi-frequency component separation, training it to segmentate the CMB from the other foreground components.

\begin{acknowledgements}
We warmly thank the anonymous referee for the very useful comments on the original manuscript.
JMC, JGN, LB, MMC and DC acknowledge financial support from the PGC 2018 project PGC2018-101948-B-I00 (MICINN, FEDER). DH acknowledges the Spanish MINECO and the Spanish Ministerio de Ciencia, Innovación y Universidades for partial financial support under project PGC2018-101814-B-I00. MMC acknowledges PAPI-20-PF-23 (Universidad de Oviedo). JDCJ, MLS, SLSG, JDS and FSL acknowledge financial support from the I+D 2017 project AYA2017-89121-P and support from the European Union’s Horizon 2020 research and innovation programme under the H2020-INFRAIA-2018-2020 grant agreement No 210489629.\\
This research has made use of the python packages \texttt{ipython} \citep{ipython}, \texttt{matplotlib} \citep{matplotlib}, \texttt{TensorFlow} \citep{tensorflow}, \texttt{Numpy} \citep{numpy} and \texttt{Scipy} \citep{scipy}, also the \texttt{HEALPix} \citep{GOR05} and \texttt{healpy} \citep{zon19} packages.
\end{acknowledgements}

%
%
\bibliographystyle{aa}
\bibliography{SDNN}

\end{document}